\documentclass[%
 reprint,
 amsmath,amssymb,
 aps,
]{revtex4-2}

\usepackage{graphicx}
\usepackage{dcolumn}
\usepackage{bm}
\usepackage{amsmath}
\usepackage{comment}
\usepackage[dvipsnames]{xcolor}
\usepackage{subfigure}

\begin{document}

\preprint{APS/123-QED}

\title{Multichannel photoelectron phase lag across atomic barium autoionizing resonances}

\author{Yimeng Wang}
\author{Chris H. Greene}%
\affiliation{Department of Physics and Astronomy, Purdue University, West Lafayette, Indiana 47907, USA and
Purdue Quantum Science and Engineering Institute,
Purdue University, West Lafayette, Indiana 47907, USA}

\date{\today}

\begin{abstract}
Phase lag associated with coherent control where an excited system decays into more than one product channel has been subjected to numerous investigations. 
Although previous theoretical studies have treated the phase lag across resonances in model calculations, quantitative agreement has never been achieved between the theoretical model and an experimental measurement of the phase lag from the $\omega-2\omega$ ionization of atomic barium, Yamazaki and Elliot, {PhysRevLett.98.053001(2007), PhysRevA.76.053401(2007)}, suggesting that a toy model with phenomenological parameters is inadequate to describe the observed phase lag behavior. 
Here the phase lag is treated quantitatively in a multichannel coupling formulation, and our calculation based on a multichannel quantum defect and $R$-matrix treatment achieves good agreement with the experimental observations. Our treatment also develops formulas to describe the effects of hyperfine depolarization on multiphoton ionization processes.  Moreover, we identify resonances between $Ba^{+}$ $5d_{3/2}$ and $5d_{5/2}$ thresholds that have apparently never been experimentally observed and classified. 
\end{abstract}

\maketitle
\section{Introduction}
The phase lag associated with coherent control by two-pathway excitation has been widely studied in both the experimental \cite{PhysRevA.76.053401,PhysRevLett.98.053001,Zhu77,PhysRevLett.92.113002,PhysRevLett.74.4799,PhysRevLett.96.173001,PhysRevLett.79.4108} and theoretical \cite{PhysRevLett.86.5454,PhysRevLett.79.4108,Nakajima_1997,PhysRevLett.82.2266,PhysRevLett.82.65,PhysRevA.50.595,Seideman:1999,PhysRevLett.70.1081} literature. 
Coherent control is based on quantum interference, in scenarios where different optical routes lead coherently to the same final state.  With a controllable optical phase difference $\Delta\varphi$ between the two laser electric fields, one can manipulate the interference between them and thereby control an observable outcome. The outcome under control shows a sinusoidal modulation of $\Delta\varphi$. i.e. for an observable $p$, $p=p_0+p_1\cos{(\Delta\varphi-\delta)}$. An especially interesting situation arises when the final state can decay into more than one continuum, and those continua respond differently to the optical phase difference. The phase lag from two continua $i,i^{\prime}$ is defined by $\Delta\delta=\delta^{(i)}-\delta^{(i^{\prime})}$, which is non-zero for most atoms and molecule.  
It has been found by many experiments that the phase lag is almost constant as a function of wavelength for flat continua, but varies rapidly at a resonance \cite{Seideman:1998,PhysRevLett.79.4108,PhysRevLett.82.65}. The phase lag behavior in the vicinity of resonances has not only drawn the attention of experimentalists interested in potential applications \cite{Zhu77,PhysRevLett.92.113002,PhysRevLett.74.4799,PhysRevLett.96.173001},  but it has also triggered a series of theoretical investigations aimed at the extraction of insights into the nature of multichannel systems. 

\begin{figure}[htbp]
  \includegraphics[scale=0.45]{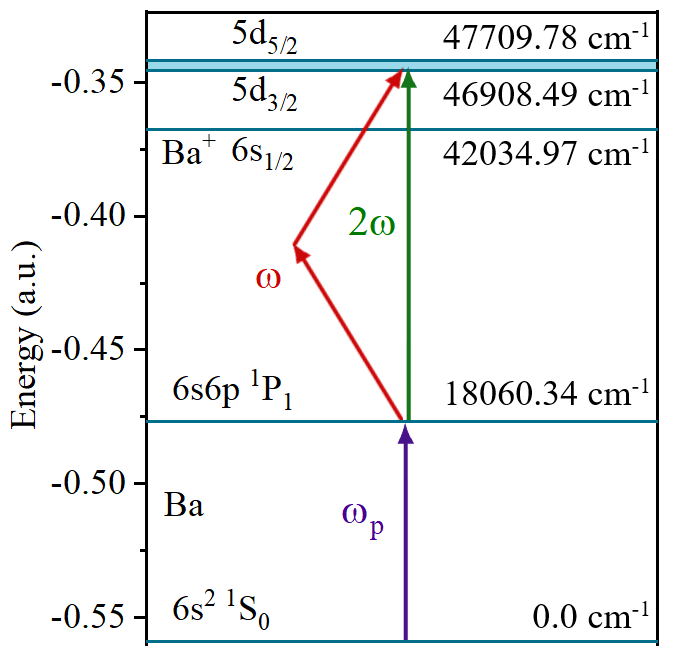}
    \caption{ Energy level diagram for the barium $\omega-2\omega$ interference scheme. Start with a ground state barium $i$, and first use a linearly-polarized pump laser ($\omega_p$) to excite the atom to the $e_1=6s6p$ $(^1P_1)$ state. Next, by the concurrent one- and two-photon ionization of that initially excited state (with fundamental frequency $\omega$), the atom reaches the shaded energy region, which can decay into either the continua associated with the $6s_{1/2}$ or the  $5d_{3/2}$ ionic state. 
    The relevant threshold energy levels in atomic unit are $E_{6s_{1/2}}=-0.3676$ $a.u.$, $E_{5d_{3/2}}=-0.3454$ $a.u.$ and $E_{5d_{5/2}}=-0.3418$ $a.u.$ (the double ionization threshold is $0$ $a.u.$), and their values in $cm^{-1}$ are given in the figure.
    }
  \label{1}
\end{figure}

Previous theoretical efforts to clarify the origin and properties of the phase lag are based on simplified models of a few discrete states embedded in two continua \cite{PhysRevLett.86.5454,PhysRevLett.79.4108,Nakajima_1997,PhysRevLett.82.2266,PhysRevLett.82.65,PhysRevA.50.595,Seideman:1999,PhysRevLett.70.1081}. However, the first quantitative comparison between that type of theoretical model and an experimental observation in the final state resonance energy range, conducted by Yamazaki et. al., was far from satisfactory \cite{PhysRevLett.98.053001,PhysRevA.76.053401}. This suggests that a toy model with phenomenological parameters is not adequate to predict the phase lag behavior in a multichannel, highly-correlated atom such as barium. In this article, we provide an {\it ab initio} calculation for the experiment of Yamazaki et. al., based on a multi-channel quantum defect (MQDT) and $R$-matrix treatment\cite{Greene:1979,Seaton:1983,GRF1982,ErratumGRF1982,Review:1996}, and obtain quantitative agreement with the experimental observations. 
The system considered by the experiment and by our calculation is the $\omega-2\omega$ concurrent ionization of atomic barium, with final state energy above two ionization thresholds, namely  $6s_{1/2}$ and $5d_{3/2}$ (Fig. \ref{1}). The role of resonances in our calculated phase lag has been analyzed from a multichannel coupling point of view. 

\section{Method}
In contrast to the previous experiments that focused on controlling chemical products of a photprocess\cite{Zhu77,PhysRevLett.92.113002,PhysRevLett.74.4799,PhysRevLett.79.4108}, the $\omega-2\omega$ scheme does not affect the total yield. 
This is because the absorption of one- or two-photons produces final states with opposite parities $\pi=\{e,o\}$, whereby the interference terms that mix different parities must have odd parity and cannot be observed by measuring the overall ionization rate.\cite{PhysRevA.103.053118,Anderson:1992,PhysRevLett.69.2353} 
The photoelectron angular distribution (PAD), which can be parameterized by symmetric ($S$) and anti-symmetric ($A$) components as $S(\theta,\phi)+A(\theta,\phi)\cos{(\Delta\varphi+\delta_0)}$ \cite{PhysRevA.76.053401}, is the observable explored here. 
The photoelectrons can escape leaving the ion in two different energy eigenstates $a$ and $b$ (referred to loosely in the following as channels), with  {principal and angular quantum numbers} $(n_c L_{c_{J_c}})$ of the ionic valence electron. For convenience, we consider one of the channels with the channel index being omitted, until later in the article where we explicitly treat the difference between the channels.
 {The differential ionization rate $\frac{dW}{d\Omega}$ can be expressed in terms of spherical harmonics $Y_{l m}$ with complex coefficients $fe^{i\delta}$ as \cite{Lindsay:1992}}:

\begin{widetext}
\begin{equation}
\label{eq-1}
\begin{split}
&\frac{dW}{d\Omega}=\mathcal{N}^2\sum_{\lambda_i}
\left| \mathcal{E}_{\omega}^2\sum_{l=odd} Y_{l m}(\theta,\phi) f_{\gamma,2}e^{i\delta_{\gamma,2}} 
+ \mathcal{E}_{2\omega} e^{i \Delta\varphi}\sum_{l=even} Y_{l m}(\theta,\phi) f_{\gamma,1}e^{i\delta_{\gamma,1}}
\right|^2\\
&f_{\gamma,q}e^{i\delta_{\gamma,q}}=\sum_{\lambda_c}
\langle J_{cs}M_{J_{cs}},l m|J_f,M_{J_f}\rangle  
\langle [(J_{c},s)J_{cs},l]J_f|[J_{c},(l,s)j]J_f\rangle T_{f,i}^{(q)}.
 \end{split}
\end{equation}
\end{widetext}
The description of atomic barium is focused on its two active electrons, with the inner and outer electron quantum numbers denoted by $(L_{c},s_c){J_c}$ and $(l,s)j$.  $\vec{J_{cs}}=\vec{J_{c}}+\vec{s}$, $\vec{J_f}=\vec{J_{cs}}+\vec{l}$.  {$\mathcal{N}^2$ is a normalization constant.} The incoherent sum index ${\lambda_i}$ includes $J_{cs}$, the nuclear spin $I$, and all the angular momentum projections $M$. The coherent sum index in Eq.\ref{eq-1} represents ${\lambda_c}=\{J_f,j\}$, and another index is defined as 
$\gamma=\{\lambda_i,l\}$. Here $T_{f,i}^{(q)}$ is the $q$-photon transition amplitude from state $i$ to state $f$, whose formula is given in the Appendix, Eq. \ref{eq2-2}. Because different pathways $q=1,2$ produce opposite parities $(-1)^l$, the index $q$ is omitted below. The two electric field strengths are $\mathcal{E}_{2\omega,\omega}$ for the one- and two-photon processes, with corresponding optical phase difference $\Delta\varphi=2\varphi_{\omega}-\varphi_{2\omega}$. 
Because both the pump and ionization lasers are linearly polarized along $\hat{z}$, the system has azimuthal symmetry. 
Eq. \ref{eq-1} can therefore be arranged into a sum of Legendre polynomials $P_k(\cos{\theta})$ with real coefficients $\beta_k$ as,
\begin{equation}
\begin{split}
 \frac{dW}{d\Omega}&=\frac{W_{tot}}{4\pi}\sum_{k=0}^{6}\beta_k P_k(\cos{\theta})\\
\end{split}    
\end{equation}
Since we have three-photon ionization at most from the ground state, $k=6$ is the maximum order. The even and odd orders of $P_{k}(\cos{\theta})$ give the symmetric and anti-symmetric contibutions to the photoelectron angular distributions(PADs). 

The degree of asymmetry of the PAD along the z-axis is quantified by the directional asymmetry parameter, defined as $\alpha_{asym}=W_{-z}/W_{tot}$, the ratio between the $-z$ directed photoelectron current and the total \cite{PhysRevLett.98.053001,PhysRevA.76.053401,PhysRevA.103.053118}. It is given by:
\begin{equation}
\label{eq-2}
\begin{split}
    \alpha_{asym}&= \frac{2\pi}{W_{\rm tot}} \int_{\frac{\pi}{2}}^{\pi} \frac{dW(\theta)}{d\Omega}\sin \theta d\theta
    =\frac{1}{2}(1+\sum_{k} \rho_{k} \beta_{k})\\
    &\equiv\frac{1}{2}+A \cos{(\Delta\varphi-\delta_0)} 
\end{split}    
\end{equation}
$\rho_k\equiv\int_{-1}^{0} P_k(x) dx$, which is $\{-\frac{1}{2},\frac{1}{8},-\frac{1}{16}\}$ for $k=\{1,3,5\}$ and is 0 when $k$ is even. 
The second line of Eq. \ref{eq-2} recasts $\alpha_{asym}$ in terms of an amplitude $A$ and phase $\delta_0$. $0\leq A\leq \frac{1}{2}$ and $0\leq\delta_0\leq2\pi$. 
The electric field strengths for the fields with frequencies $\mathcal{E}_{\omega,2\omega}$ affect the amplitude $A$ only. The phase $\delta_0$ can be expressed in terms of $fe^{i\delta}$ and angular momentum coefficients as: 
\begin{equation}
\label{eq-3}
\begin{split}
    \delta_0&=-arg\left[\sum_{\lambda_i,l_o,l_e,k_o}\rho_{k_o}\Theta(l_o,l_e,k_o,m) f_{\gamma_o}f_{\gamma_e}e^{i(\delta_{\gamma_e}-\delta_{\gamma_o})}\right]
\end{split}    
\end{equation}
where 
\begin{small}
\begin{equation*}
\begin{split}
\Theta(l,l^{\prime},k,m)&= (2k+1)\int Y_{l,m}(\theta,\phi) P_{k}(\cos{\theta}) Y_{l^{\prime},m}^{*}(\theta,\phi) d\Omega\\
\end{split}    
\end{equation*}
\end{small}
Note that the subscript $e$($o$) for $l$ and $k$ denotes even(odd) numbers, and $\gamma_{e(o)}=\{\lambda_i,l_{e(o)}\}$. $\delta_0$ comes from the interference terms between $l_{e}$ and $l_{o}$.  
The phase lag is defined as $\Delta\delta\equiv \delta_0^{(a)}-\delta_0^{(b)}$. For the ionization scheme being considered for the Ba atom, the two channels are $a=6s_{1/2}$ and $b=5d_{3/2}$. The energetically closed $5d_{5/2}$ channel supports autoionizing states. Angular momenta correspond to those channels are listed in Table.\ref{tab:table1}. 
The energy-dependent calculations carried out here are plotted versus final state energies  {from 47200 to 47265 $cm^{-1}$ relative to the ground state}, which is roughly one cycle of the $5d_{5/2}$ Rydberg series, with the effective principle quantum number of the fragmentation electron being $\nu=1/[2(E_{5d_{5/2}}-E)]^{\frac{1}{2}}=14.7-15.7$.
Barium photoionization spectra in this energy range have been previously measured \cite{PhysRevLett.98.053001,PhysRevA.76.053401,Camus_1983,Maeda_2000,Aymar_1990} in the laser wavelength range $\lambda_{\omega}= 684.82 - 686.35$ $nm$. 


\begin{table}[htbp]
\centering
\caption{\label{tab:table1} Channels relevant to evaluating equations \ref{eq-1}-\ref{eq-3}, represented by total angular momentum and parity $J_f^{\,\pi}$ and jj - coupled basis $n_c L_{c}{J_c} (l, j)$  {(the subscript $c$ denotes ion core)}. The row ``channel $a (b)$" lists the angular momenta for the fragmentation electron. The row ``resonance" gives the angular momenta of the autoionizing states that converge to the $5d_{5/2}$ threshold. }
\begin{footnotesize}
\begin{tabular}{c|ccc|ccc}
 \hline
 \hline
 &\multicolumn{3}{c|}{one-photon paths}&\multicolumn{3}{c}{two-photon paths}\\
 $J_f^{\,\pi}$& $\quad0^{e}\quad$ & $\quad1^{e}\quad$& $\quad2^{e}\quad$& $\quad1^{o}\quad$& $\quad2^{o}\quad$ & $\quad3^{o}\quad$    \\
 \hline
 $6s_{1/2}\epsilon (l,j)$ & $(0,1/2)$ & $(0,1/2)$& $(2,3/2)$& $(1,1/2)$& $(1,3/2)$ & $(3,5/2)$ \\
 (channel $a$)&  & $(2,3/2)$ & $(2,5/2)$& $(1,3/2)$& $(3,5/2)$ & $(3,7/2)$  \\  
 \hline
 $5d_{3/2}\epsilon (l,j)$ & $(2,3/2)$ & $(0,1/2)$& $(0,1/2)$& $(1,1/2)$& $(1,1/2)$ & $(1,3/2)$ \\
 (channel $b$)&  & $(2,3/2)$& $(2,3/2)$& $(1,3/2)$& $(1,3/2)$ & $(3,5/2)$  \\ 
  &  &$(2,5/2)$& $(2,5/2)$& $(3,5/2)$& $(3,5/2)$ & $(3,7/2)$  \\ 
  &  &  & $(4,7/2)$&  & $(3,7/2)$ &    \\ 
 \hline
 \hline 
 $5d_{5/2}n (l,j)$& $(2,5/2)$ & $(2,3/2)$& $(0,1/2)$& $(1,3/2)$& $(1,1/2)$ & $(1,1/2)$ \\
 (resonances)&  & $(2,5/2)$& $(2,3/2)$& $(3,5/2)$& $(1,3/2)$ & $(1,3/2)$ \\
  &  & $(4,7/2)$& $(2,5/2)$& $(3,7/2)$& $(3,5/2)$ & $(3,5/2)$ \\
  &  &  & $(4,7/2)$&  & $(3,7/2)$ & $(3,7/2)$ \\
  &  &  & $(4,9/2)$&  &  &  \\
 \hline
 \hline  
\end{tabular}
\end{footnotesize}
\end{table}

\begin{figure}[htbp]
\centering
    \begin{subfigure}
        \centering
        \includegraphics[scale=0.44]{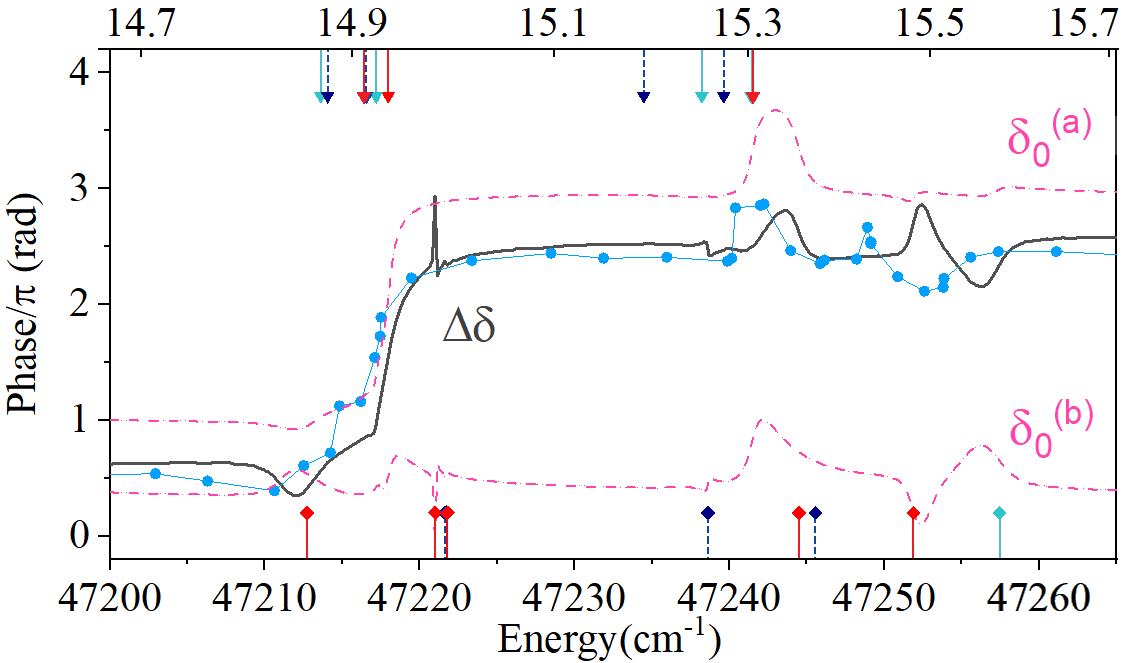}
        \caption{Comparison of our calculated phase lag $\Delta\delta=\delta_0^{(a)}-\delta_0^{(b)}$ (solid curve) with the experimental results  \cite{PhysRevLett.98.053001} (blue points) versus energy in $cm^{-1}$.  {The top label gives the effective quantum number $\nu=[2(E_{5d_{5/2}}-E)]^{-\frac{1}{2}}$.} 
        The dashed curves are $\delta_0$ for $a=6s_{1/2}$ (upper) and $b=5d_{3/2}$ (lower) channels. The arrows and the diamonds on the top and bottom give the positions of the two- and one-photon resonances, and their colors correspond to the colors of the different partial waves $J_f$ shown in Fig. \ref{3}. }
        \label{2}
    \end{subfigure}
\hfill
    \begin{subfigure}
        \centering
        \includegraphics[scale=0.36]{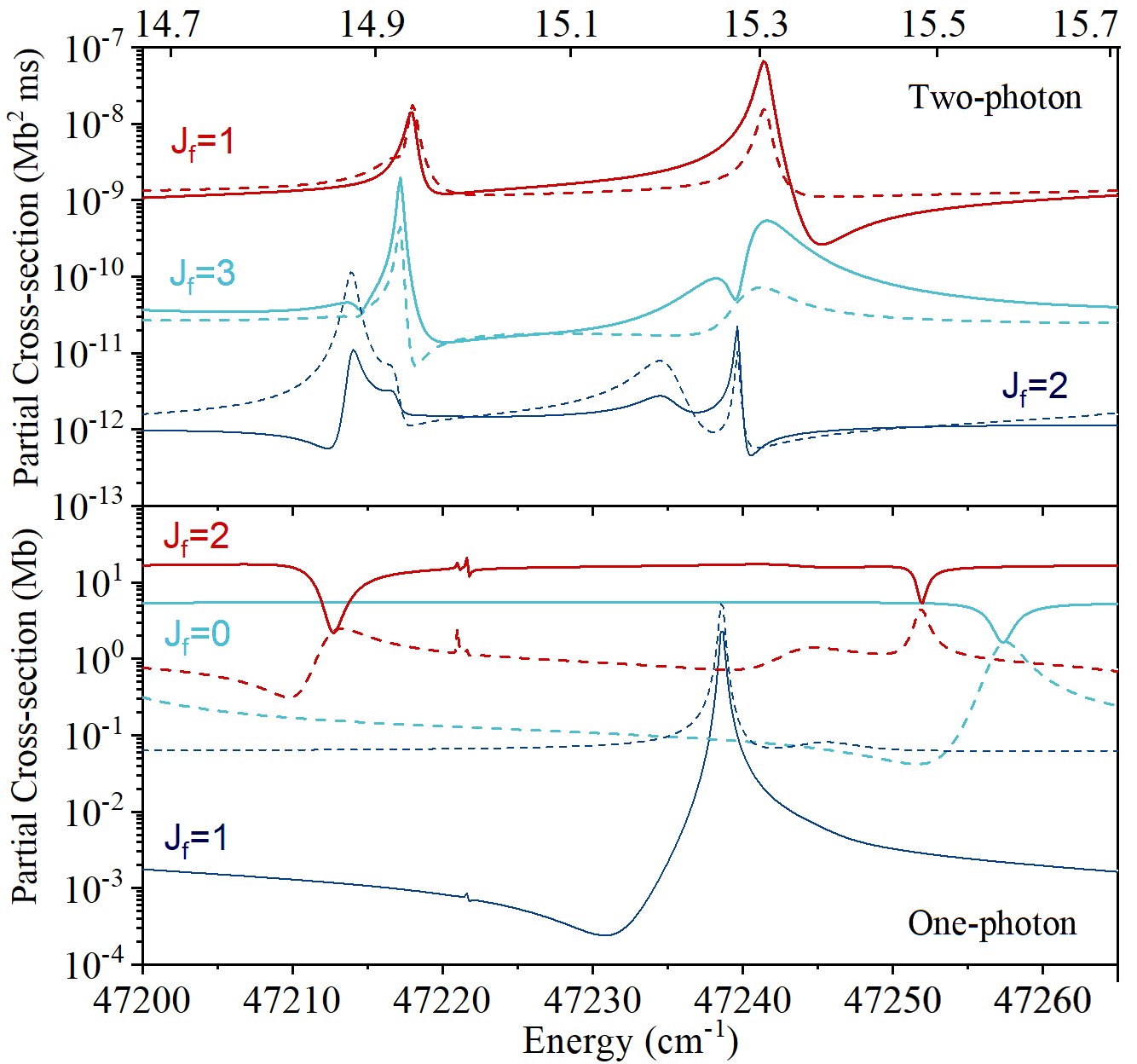}
        \caption{Partial cross-sections for two- (upper panel) and one-photon (lower panel) ionization, separately. 
        The solid (dashed) curves give partial cross sections for $a=6s_{1/2}$($b=5d_{3/2}$) channel. In both figures, the parity unfavored partial waves enabled by hyperfine depolarization have the smallest component.   
Resonance identification and their comparisons with experiment are given in Table. \ref{tab:table3}.}
        \label{3}
    \end{subfigure}
\end{figure}

Previous theoretical investigations of the phase $\delta_0$ were based on the partitioned Lippmann-Schwinger equation, which separates the transition amplitude $fe^{i\delta}$ into a direct ($f^d e^{i\delta^d}$) and a resonance-mediated ($f^r e^{i\delta^r}$) component \cite{PhysRevLett.79.4108,PhysRevLett.82.65,Nakajima_1997,PhysRevLett.82.2266,PhysRevA.76.053401}. The first term describes the direct ionization from the initial state $i$ directly to the continuum final state $\xi$, while the latter describes the pathway to the continuum via an autoionizing state $n$ that couples to $\xi$ through an electron correlation matrix element $V_{n \xi}$. These two amplitudes are represented as: 
\begin{equation}
\label{eq-4}
f^d e^{i\delta^d}=D_{\xi i} \qquad f^r e^{i\delta^r}=\sum_n \frac{V_{\xi n}\Omega_{ni}(1-{i}/{q_{n}})}{E-(E_n-i\Gamma_n/2)},
\end{equation}
where $\Omega_{ni}(1-\frac{i}{q_{n}})=D_{ni}+\sum_{\xi^{\prime}}\int dE_{\xi^{\prime}} \frac{V_{n \xi^{\prime}}D_{\xi^{\prime} i}}{E_n-E_{\xi^{\prime}}}$, $q_{n}$ is the Fano lineshape asymmetry parameter and $\Gamma_n$ is the line width of resonance $n$. 
This model attributes the large value of $\Delta\delta$ and its strong variation near the resonance to the interference between $f^d e^{i\delta^d}$ and $f^r e^{i\delta^r}$. When $f^r e^{i\delta^r}=0$ (far from resonances), $\delta_0(E)$ is almost constant. When $D_{\xi i}=0$, $q_{n}\rightarrow\infty$ (in the centre of an isolated symmetric resonance), $\delta_0$ are determined by the complex resonance energy $(E_n-i\Gamma_n/2)$, in situations discussed by those papers where both optical paths lead to the same final state,  $\Delta\delta$ obtain a local minimum. 
The Yamazaki {\it et al.} theory comparison with their experiment used Eq. \ref{eq-4} to extract resonance parameters $q_{n}, \Gamma_n$ by fitting to their spectra, and it showed a large discrepancy. 
One sees, in Fig. \ref{2} for example, where the experimentally observed $\Delta\delta$ is represented by the blue points, a huge phase variation of $2\pi$ is observed at $J_f^{\pi}=1^{o}$ resonances, while Yamazaki {\it et al.} predicts a maximum phase lag about $\pi/3$ and $7\pi/6$ due to the resonance effect and the outgoing waves  {(For their theoretical results, see Ref.\cite{PhysRevA.76.053401} Fig. 11 and 12)}.

Our analysis of the phase lag $\Delta\delta$ derives from a different picture than the aforementioned single-resonance model, as ours is based on the  multichannel quantum defect (MQDT) \cite{Greene:1979,Seaton:1983,GRF1982,ErratumGRF1982} and the streamlined $R$-matrix method \cite{Review:1996}.   We assume that the electron correlations are restricted to the region where both of the electrons are inside a reaction zone (i.e. in this calculation, within 60 $a.u.$ from the nucleus). 
In contrast to the traditional Lippmann-Schwinger model that considered the coupling only between a few autoionizing and continuum states, our $R$-matrix treatment describes the coupling between all the two-electron basis functions $\mathcal{Y}(n_c L_{c_{J_c}};nl_j)$ inside the reaction zone, with $L_c,l = 0-4$, and up to 60 nodes in the  radial basis functions (For more details, refer to Ref. \cite{PhysRevA.80.033401,PhysRevA.103.033103}). The electron correlation is therefore modeled in a realistic manner, and the resonance structure is described in its full multichannel complexity.

When one of the electron moves beyond the reaction zone, it experiences an overall Coulomb potential from the other part of the system. The energy range treated here is $E_{6s_{1/2}},E_{5d_{3/2}}<E<E_{5d_{5/2}}$. Depending on whether the inner electron resides in the $6s_{1/2}$, $5d_{3/2}$ or $5d_{5/2}$ state, it can be described by Coulomb functions with either incoming-wave boundary conditions appropriate to a photofragmentation process into the $E_{6s_{1/2}},E_{5d_{3/2}}$ channels \cite{PhysRev.93.888} or an exponentially decaying boundary condition ($5d_{5/2}$ and higher channels) \cite{Greene:1985}. 
The asymptotic wave function of barium can be written in terms of the superposition of those channel states as, 
\begin{small}
\begin{equation*}
\begin{split}
      \psi_{i^{\prime}}\rightarrow&
      {\cal A} \sum_{i\in o}\frac{\Phi_i}{ir}\frac{1}{\sqrt{2\pi k_i}}(e^{ik_ir}r^{i/k_i}\delta_{ii^{\prime}}-e^{-ik_ir}r^{-i/k_i}{S}^{\dagger phys}_{ii^{\prime}}) \\
      +&{\cal A} \sum_{i\in c}\frac{\Phi_i}{r} W_i(r) Z_{ii^{\prime}}%
\end{split}
\end{equation*}
\end{small}
The channel function $\Phi_i$ includes all the other parts of the barium except the radial motion of the fragmentation electron. $i=\{n_c L_{c_{J_c}},l_j\}$. 
$i\in o$ or $c$ indicates  {the quantum numbers of ion core} $n_c L_{c_{J_c}}=6s_{1/2},5d_{3/2}$ or $5d_{5/2}$. $\underline{S}^{phys}$ is the physical scattering matrix and $\underline{Z}_{co}$ is the coefficient of the exponentially decaying (rescaled) Whittaker function $W_i(r)$: 
\begin{equation}
\label{eq-5}
\begin{split}
\underline{S}^{\dagger phys}&=e^{-i\underline{\eta}}\underline{\tilde{S}}^{\dagger}e^{-i\underline{\eta}} \qquad
\underline{\tilde{S}}^{\dagger}= \underline{S}^{\dagger}_{oo}-\underline{S}^{\dagger}_{oc}(\underline{S}^{\dagger}_{cc}-e^{2i\underline{\beta}})^{-1}\underline{S}^{\dagger}_{co} \\
\underline{Z}_{co}&= e^{i\underline{\beta}}(\underline{S}^{\dagger}_{cc}-e^{2i\underline{\beta}})^{-1}\underline{S}^{\dagger}_{co}e^{-i\underline{\eta}} \\
\end{split}
\end{equation}
${\eta}$ is the Coulomb plus centrifugal phase, and is almost constant across this energy range, and the negative energy phase parameter is ${\beta}=\pi(\nu-l)$. 
Considering the ionization from an initial state $i$ directly to $\psi_{i^{\prime}}$, the components with coefficients 
$\underline{S}^{\dagger}_{oc}(\underline{S}^{\dagger}_{cc}-e^{2i\underline{\beta}})^{-1}\underline{S}^{\dagger}_{co}$ and $\underline{Z}_{co}$
which represent a pathway through the autoionizing states, correspond to $f^r e^{i\delta^r}$; and components with coefficients $\underline{S}^{\dagger}_{oo}$, which include the direct transition and transition through couplings between the continua, correspond to $f^d e^{i\delta^d}$.  The eigenvalues of $\underline{\tilde{S}}^{\dagger}$ are $e^{-2i\pi\tau_{\rho}}$\cite{Greene:1985}, as the energy increases across a single isolated resonance, the sum of eigenphases  $\sum_{\rho}\tau_{\rho}$ changes by 1, which is fundamentally the origin of the variation of the phase lag.  
The significant differences between our calculated results and the simplified resonance model implemented in Ref.\cite{PhysRevA.76.053401}  suggests that electron correlation effects, and the rich number of multiple interacting resonances in barium, plays a crucial role in this energy range of barium. 

\section{Results and Discussion}
\begin{table*}
\caption{\label{tab:table3}The resonances from our calculation (position $E$ and width $\Gamma$) and earlier experiments (Ref. \cite{Camus_1983,Maeda_2000}). Most experimental two-photon resonances are missing, because of the strong dominance of $J_f=1$ partial wave. Our classification of the resonances is shown in the $J_f(nl_{j})$ basis. For those overlapped resonances, the upper index of $\Gamma$ give the $J_f$.  }
\begin{ruledtabular}
\begin{tabular}{c|cccccccccc}
\multicolumn{11}{c}{identification of even-parity resonances $5d_{5/2}nl_{j}$ }\\
 \hline
 $\nu$    &14.86 & 14.98 &14.99  &15.26 &  15.36 &15.38 & 15.48 & 15.58& \\ 
$E(cm^{-1})$ & 47212.77 & 47220.99  & 47221.62 & 47238.53 & 47244.70 & 47245.65 & 47251.81 &47257.50& &\\
$\Gamma(cm^{-1})$ & 3.29  &  0.06 & 0.07$^{1}$,0.06$^{2}$  & 0.42 & 5.74 & 5.47 & 1.15 & 4.63 & &\\
 \hline 
 Exp \cite{Camus_1983} & 47210.86  & 47220.59 &  47221.18& 47238.77 & 47241.75  & 47244.56 & 47250.81 &47256.77& &\\
 \hline 
  $J_f^{\,e}(nl_{j})$&$2(19s_{1/2})$&$2(15g_{7/2})$&$1(15g_{7/2})$&$1(18d_{5/2})$&$2(18d_{3/2})$&$1(18d_{3/2})$&$2(18d_{5/2})$&$0(18d_{5/2})$& &\\
  & & &$2(15g_{9/2})$& & & & & & & \\  
 \hline
 \hline 
\multicolumn{11}{c}{identification of odd-parity resonances $5d_{5/2}nl_{j}$}\\
 \hline
 $\nu$ & 14.87 & 14.88 & 14.91 &14.92 & 14.93 & 14.94 & 15.19 & 15.26 & 15.28 & 15.31 \\
 $E(cm^{-1})$ & 47213.61 & 47214.05 &47216.39 & 47216.55& 47217.18& 47217.99 & 47234.47 & 47238.23 & 47239.63 &47241.4  \\
 $\Gamma(cm^{-1})$ & 1.34 & 1.02 & 2.16 & 1.11 & 0.40 & 1.08  & 3.32 & 2.69 & 0.41 &1.0$^{1}$,3.1$^{3}$   \\ 
 \hline 
 Exp \cite{Maeda_2000} &  & & &  &  & 47218.27&  &  &  &  47241.10    \\
 \hline 
$J_f^{\,o}(nl_{j})$&$3(15f_{5/2})$&$2(15f_{7/2})$&$1(15f_{5/2})$&$2(15f_{5/2})$&$3(15f_{7/2})$&$1(15f_{7/2})$&$2(19p_{1/2})$&$3(19p_{1/2})$&$2(19p_{3/2})$&$1(19p_{3/2})$ \\ 
 & & &&& & & & & &$3(19p_{3/2})$ \\ 
\end{tabular}
\end{ruledtabular}
\end{table*}

Our computed phase lag and its experimental observations 
are presented in Fig. \ref{2}, shown respectively as the solid curve and the points.  {When $\nu<15.1$}, the two results agrees well, but a significant energy shift of features of around $1.6\times 10^{-5}$ $a.u.$ shows up when $\nu>15.1$, which is larger than the difference between the theoretical and experimental resonance positions listed in Table. \ref{tab:table3}. The calculated resonance positions are indicated by the diamonds and the arrows. The ones on the bottom (top) are for one (two)-photon pathway autoionization, and the color labels the resonance angular momentum $J_f$. The dashed curves in Fig. \ref{2} give the interference phases $\delta_0^{(a),(b)}$ for each channel; after each resonance-caused variation, they return to their initial values at $\delta_0^{(a)}=\pi$ and $\delta_0^{(b)}=0.38\pi$. This indicates that the phase change across the resonances either shift by $2N\pi$, or else they shift back and forth by an arbitrary value, returning to their initial value.

\begin{figure}[htbp]
  \includegraphics[scale=0.32]{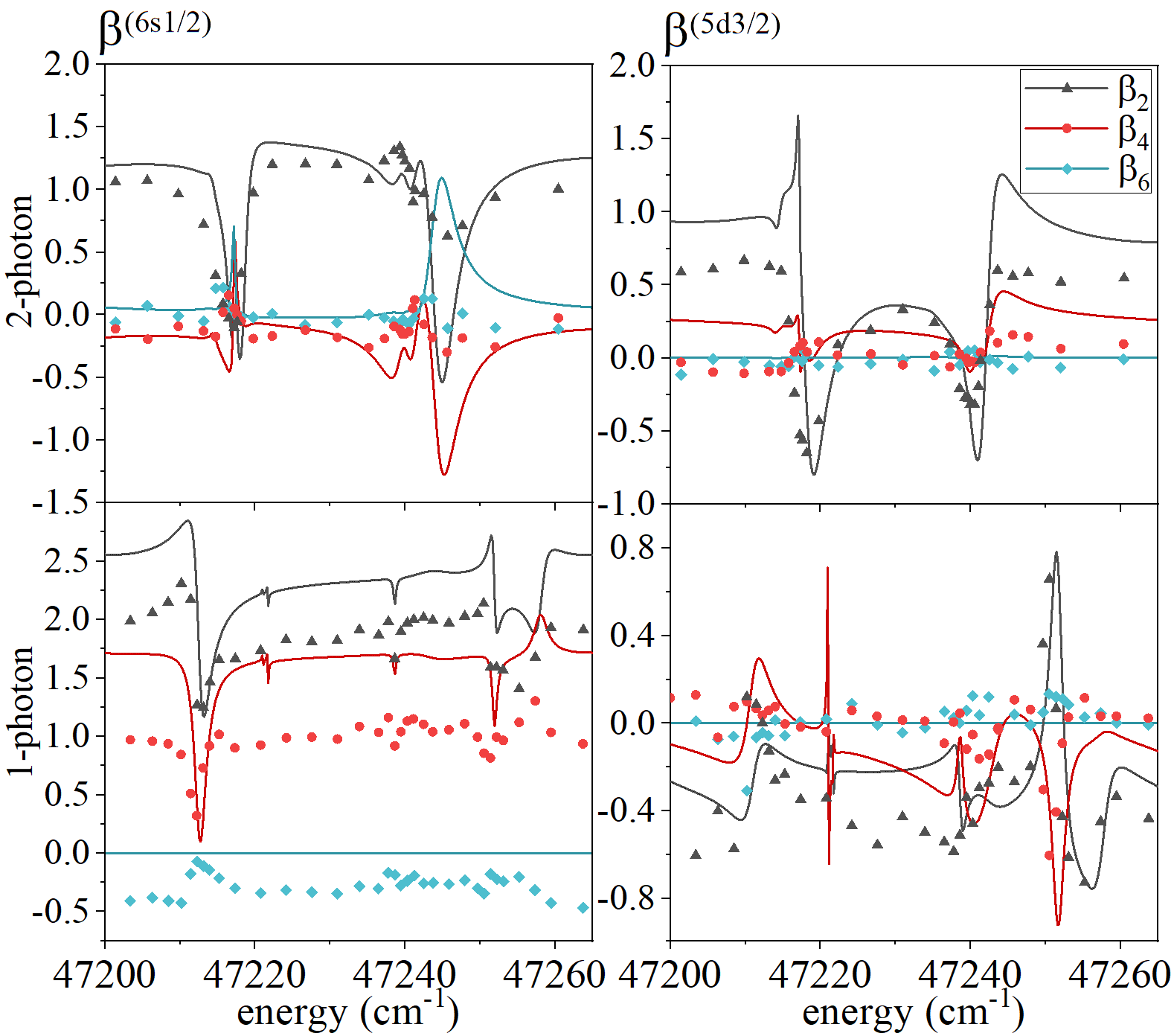}
    \caption{ The photoelectron angular distribution parameters $\beta_k^{(a,b)}$ for one- and two-photon ionization process. Without interference effect, the odd-order parameters $\beta_{k_o}=0$. Our calculations (solid curves) are compared to the experimental results (points) from Ref. \cite{PhysRevA.76.053401} Fig. 6 and 7. 
    The upper and lower panels are the $\beta$ parameters for two- and one-photon ionizations respectively, and the left and right panels are for channels $(a)$ and $(b)$. }
  \label{4}
\end{figure}

To analyze the influence of resonances to the phase shifts,  Fig. \ref{3} presents our calculated partial cross-sections (Eq. \ref{eq2-3}) for one- and two-photon processes separately. 
Among the two-photon pathways, the $J_f=1$ partial wave dominate over other partial wave by orders of magnitude. The $J_f=3$ resonances, often overlapping with stronger $J_f=1$ resonances, are therefore hard to observe. 
Since $\delta_0$ is obtained by a coherent sum over all the partial waves (Eq. \ref{eq-3}), those with tiny amplitudes do not play a major role. Consequently, only the three resonances from $J_f=1$ (the red arrows in Fig. \ref{2}) can appreciably influence the interference phase.  
Similarly, among the one-photon pathways, it is the resonances from the prominent $J_f=2$ partial wave, and $J_f=$1, 0 resonances at  {$E=$ 47238.53 $cm^{-1}$ and 47257.50 $cm^{-1}$} whose signals are strong enough to affect $\delta_0$. Besides, the influence of those resonances to different channels appears quite differently: $\delta_0^{(b)}$ responds sensitively to both one- and two-photon resonances, while $\delta_0^{(a)}$ responds strongly to the two-photon resonances, but remains almost unaltered by one-photon resonances. However, we cannot provide a straightforward model to explain why the one- and two-photon resonances can influence $6s_{1/2}$ and $5d_{3/2}$ channels in such a completely difference manner. 

Besides the comparison in phase lag, we calculated the  $\beta_k^{(a,b)}$ for separate one- and two-photon pathways in Fig. \ref{4}. The threshold energies we used in this calculation are calibrated according to Ref. \cite{PhysRevA.76.053401}, to compare with its experimental results. The calculated $\beta_k$ (solid lines) are in general larger than the experimental ones (points), but similarities can still be found in their lineshapes. Noted that $\beta_6$ for one-photon pathway should be 0 from theoretical considerations, but the experimental fitting gave a nonzero value. In all, the agreement in the photoelectron angular distributions is less satisfactory than the phase lag results, which suggests that the phase lag could be more robust against small calculated phase and amplitude errors. 

\section{Conclusion}
To conclude, we provide quantitative agreement between theory and experiment about phase lag from optical control of photoelectron angular distribution in atomic barium. A full treatment of nonperturbative electron correlations is crucial to incorporate, in order to correctly predict the phase lag behavior, which is beyond the description of a toy-model. 
We also provide a systematic analysis of the role of hyperfine depolarization on barium multiphoton ionization, and develop formulas for differential cross-sections and three-photon ionization cross-sections, which have not been included in previous work;  this development is presented in the Appendix. 
Finally, this work provides a detailed study of the barium photoabsorption spectrum between the $5d_{3/2}$ and $5d_{5/2}$ thresholds, it identifies some resonances that are hardly resolved experimentally, and based on that it enables an attribution of the different influences of those resonances to the phases.

\begin{acknowledgments}
We thank D. S. Elliott for useful discussions. 
This work was supported by the U.S. Department of Energy, Office of Science, Basic Energy Sciences, under Award No. DE-SC0010545.
\end{acknowledgments}

\appendix*
\section{\label{sec:hfp} Hyperfine Depolarization Effects}

This Appendix develops a detailed analysis of the role of hyperfine depolarization on the photoionization process, including a derived formula for ionization amplitudes $T_{f,i}$ and cross sections under their influence. Our method of evaluating ionization amplitudes is based on time-dependent perturbation theory, with the light-atom interactions being simplified by the electric dipole approximation, as in Ref. \cite{PhysRevA.103.033103}.  

Hyperfine structure is important in atomic barium, because it breaks ordinary electronic selection rules . 
Natural barium consists of two isotopes with different nuclear spins: The $I=0$ isotopes constitute $82\%$ while the remaining $18\%$ are $I=\frac{3}{2}$ isotopes \cite{Elements}. 
The latter experience hyperfine splittings, and hence one can visualize semiclassically that the nuclear spin $\vec{I}$ and electronic angular momentum $\vec{J}$ precess about the total angular momentum $\vec{F}$. 
Differences between the hyperfine levels $E_F$ usually are in the range of hundreds of $MHz$ to a few $GHz$. 
A crucial point is that, in the first step of the experiment, where a high-frequency pump laser $\omega_p$ is used to excite ground state barium to $e_1=6s6p$ $(^1P_1)$, the resolution of the pump laser is around 1 $cm^{-1}$ which cannot resolve the hyperfine levels. Moreover, the excitation step can be viewed as {\it sudden} in comparison with the time scale of hyperfine-induced precession of $\vec{J}$.  Hence``quantum beats" between different $(F,M_F)$ are expected, with the transition amplitude $T_{f,i}$ now written as a coherent sum over the indistinguishable pathways: \cite{PhysRevA.47.229}. 
\begin{widetext}
\begin{equation}
\label{eq2-2}
\begin{split}
T_{f,i}^{(q,I=0)}=&T_{f,e_1}^{(q)} \langle J_{e_1} M_{J_{e_1}}|\hat{\mathbf{\epsilon}}_p\cdot\vec{r}_p|J_i M_{J_i}\rangle\\
T_{f,i}^{(q,hfp)}
=&T_{f,e_1}^{(q)}\sum_{F M_F M_{F_i}} \frac{e^{i\omega_{FF_i}t}}{2I+1}\langle J_{e_1} M_{J_{e_1}},I M_I^{\prime}|F,M_F\rangle 
\langle (J_{e_1},I)F,M_F|\hat{\mathbf{\epsilon}}_p\cdot\vec{r}_p|(J_i,I)F_i,M_{F_i}\rangle 
\langle F_i,M_{F_i}|J_i M_{J_i},I M_I\rangle \\
\end{split}   
\end{equation}
\end{widetext}

Where $T_{f,i}^{(q,I=0)}$ is the transition amplitude assuming no hyperfine structures, and $T_{f,i}^{(q,hfp)}$ includes the hyperfine structures and quantum beats from the laser-pump step, it depend on the orientation of $\vec{I}$. 
After averaging (summing) over initial states $I,M_I$ (final states $I,M_I^{\prime}$) in Eq. \ref{eq-1}, the whole process is azimuthal symmetric. 
Here we separate the pump step $i\rightarrow e_1$ and the ionization steps $e_1\rightarrow f$, and we neglect the precession effects during the very short time of the ionization processes $e_1\rightarrow f$.  The amplitudes for the individual ionization steps are, 

\begin{equation*}
T_{f,e_1}^{(q)}=
\begin{cases}
 \langle [J_{c},(l,s)j]J_f M_f|\hat{\mathbf{\epsilon}}_1\cdot\vec{r}_1|J_{e_1} M_{e_1}\rangle& q=1 \\ 
 &\\
 \sum_{e_2}\frac{\langle [J_{c},(l,s)j]J_f M_f|\hat{\mathbf{\epsilon}}_2\cdot\vec{r}_2|e_2\rangle
 \langle e_2|\hat{\mathbf{\epsilon}}_1\cdot\vec{r}_1|J_{e_1} M_{e_1}\rangle}{E_{e_2}-E_{e_1}-\omega}& q=2   \\
\end{cases}
\end{equation*}
The next step is to calculate the differential ionization rate $\frac{dW}{d\Omega}$ in Eq. \ref{eq-1} and the ionization cross-sections with $T_{f,i}^{(q,hfp)}$. The cross-section including both the pump step and the $q$-photon ionization step are:
\begin{equation}
\sigma_{i\rightarrow f}^{(q)}=2\pi(2\pi\alpha)^{q+1}\omega^q\omega_p\left|T_{f,i}^{(q)}\right|^2
\end{equation}

The effect of hyperfine depolarization to cross-sections can be parameterized by a factor $g^{(k)}(t)$, as stated in Ref. \cite{PhysRevA.47.229,Fano:1973} for the pump plus one-photon ionization. We generalized it into a pump plus two-photon ionization case  \cite{Zare:1983,Zare:1986}. The formula for $\sigma_{i\rightarrow f}^{(q=1,2,hfp)}$ can be expressed in terms of Wigner 6J, 9J operators, polarization tensors $[\hat{\mathbf{\epsilon}}\, \hat{\mathbf{\epsilon}}^{*}]^{(k)}_{\mu}$
(which is $E_{\mu}^{k}(\hat{\mathbf{\epsilon}},\hat{\mathbf{\epsilon}}^{*})$ in Ref. \cite{PhysRevA.47.229})
and reduced matrix elements as following ($[k]\equiv\sqrt{2k+1}$): 

\begin{widetext}
\begin{equation}
\label{eq2-1}
\begin{split}
\sigma_{i\rightarrow f}^{(1,hfp)}
=&2\pi(2\pi\alpha)^2\omega\omega_p
  \left|\langle f||r_1^{(1)}||e_1\rangle\langle e_1||r_p^{(1)}||i\rangle\right|^2\times  \sum_{k}
\sum_{\mu}(-1)^{\mu}[\hat{\mathbf{\epsilon}}_p \hat{\mathbf{\epsilon}}_p^{*}]^{(k)}_{\mu}
[\hat{\mathbf{\epsilon}}_1 \hat{\mathbf{\epsilon}}_1^{*}]^{(k)}_{-\mu} g^{(k)}(t) (-1)^{k}
    \left\{
    \begin{matrix}
    J_{e_1} & J_{e_1} & k \\
    1 & 1 & J_i
    \end{matrix}
    \right\}
    \left\{
    \begin{matrix}
    J_{e_1} & J_{e_1} & k \\
    1 & 1 & J_f
    \end{matrix}
    \right\} \\
\sigma_{i\rightarrow f}^{(2,hfp)}
=&2\pi(2\pi\alpha)^3\omega^2\omega_p\sum_{J_{e_2},J_{e_2}^{\prime}}
  \left|\langle f||r_2^{(1)}||e_2\rangle\langle e_2||r_1^{(1)}||e_1\rangle
  \langle e_1||r_p^{(1)}||i\rangle\right|^2\times 
  \sum_{k,k^{\prime},k^{\prime\prime}}
\sum_{\mu}(-1)^{\mu}[\hat{\mathbf{\epsilon}}_p \hat{\mathbf{\epsilon}}_p^{*}]^{(k)}_{\mu}
\left\{ [\hat{\mathbf{\epsilon}}_1 \hat{\mathbf{\epsilon}}_1^{*}]^{(k^{\prime})} [\hat{\mathbf{\epsilon}}_2 \hat{\mathbf{\epsilon}}_2^{*}]^{(k^{\prime\prime})}
\right\}^{(k)}_{-\mu}
  \\
 &\times g^{(k)}(t) (-1)^{k+k^{\prime}+J_i+J_{e_1}+J_{e_2}+J_f}[k^{\prime}][k^{\prime\prime}] 
    \left\{
    \begin{matrix}
    J_{e_1} & J_{e_1} & k \\
    1 & 1 & J_i
    \end{matrix}
    \right\}
    \left\{
    \begin{matrix}
    J_{e_2} & J_{e_2}^{\prime} & k^{\prime\prime} \\
    1 & 1 & J_f
    \end{matrix}
    \right\}    
    \left\{
    \begin{matrix}
    J_{e_2} & 1& J_{e_1} \\
    J_{e_2}^{\prime} & 1& J_{e_1} \\    
    k^{\prime\prime}& k^{\prime} & k
    \end{matrix}
    \right\}\\    
\end{split}
\end{equation}
\end{widetext}

where $k$ is the rank of tensor ($k=0,1,2$), and $g^{(k)}(t)$ is,
\begin{equation*}
g^{(k)}(t)=\sum_{F,F^{\prime}} \frac{(2F+1)(2F^{\prime}+1)}{2I+1} \cos{(\omega_{FF^{\prime}}t)}
    \left\{
    \begin{matrix}
    J_{e_1} & J_{e_1} & k\\
    F & F^{\prime} & I
    \end{matrix}
    \right\}^2   
\end{equation*}
$g^{(0)}(t)=1$, when $I=0$, $g^{(k)}(t)=1$. Similarly, in the limit $t\rightarrow 0$ when there is no time for spin procession, $g^{(k)}(0)=1$, and $\sigma_{i\rightarrow f}^{(q,hfp)}=\sigma_{i\rightarrow f}^{(q,I=0)}$. 
In the limit of complete depolarization $t\rightarrow \infty$, which is close to the case we studied, the terms with $F\neq F^{\prime}$ average to 0. 

\begin{figure}[htbp]
  \includegraphics[scale=0.4]{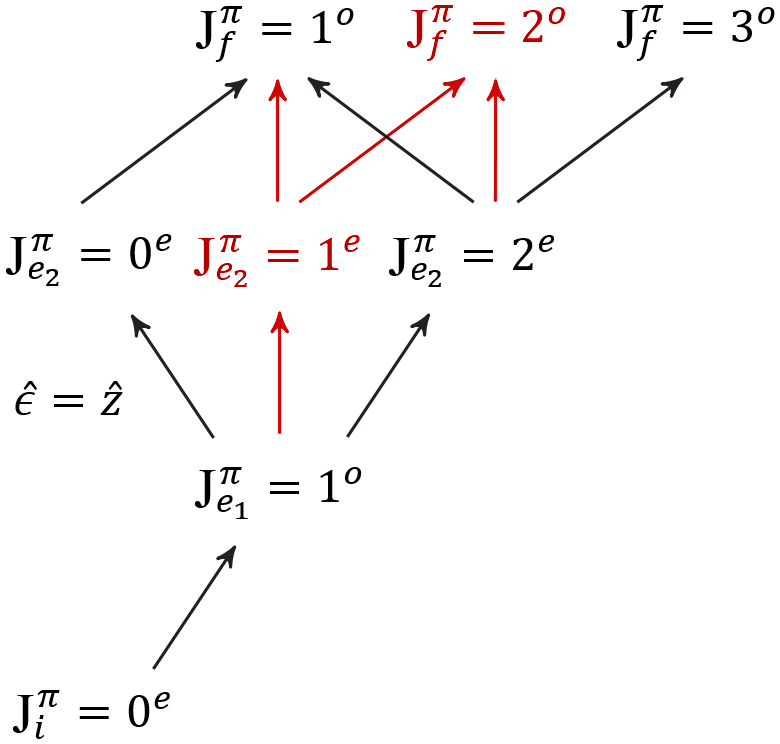}
    \caption{ The angular momenta $J$ and parity $\pi$ allowed by the electron-dipole selection rule (For the single photon $q=1$ pathway, replace $J_{e_2}$ by $J_f$). When the pump and ionization photons are both polarized along z-axis, the red paths are allowed only when the effect from hyperfine depolarization is included in the initially excited $e_1$ state. }
  \label{5}
\end{figure}

Now consider the situation in the Yamazaki et al. experiment, where the pump and ionization lasers are linearly polarized and aligned with $\hat{z}$. The relevant angular momenta $J$ for all steps are shown in Fig. \ref{5}, where the red paths are present only because of hyperfine depolarization. 
The cross-section calculations for one- and two-photon ionization ($e_1\rightarrow f$) starts from $e_1=6s6p$ $(^1P_1)$; fundamentally,  hyperfine depolarization simply mixes different $M_{J_{e_1}}$ sublevels.  To an excellent approximation, only $M_{J_{e_1}}=0$ is initially excited since all laser linear polarization axes are parallel to the $z$-axis.  Then when  ${\overrightarrow J}$ precesses about ${\overrightarrow F}$, other $M_{J_{e_1}}$ states get populated as time evolves. The ionization cross-section is determined by, 

\begin{equation}
\label{eq2-3}
\begin{split}
   &\sigma_{e_1\rightarrow f}^{(q,hfp)}%
   =\sum_{M_{J_{e_1}}}\sigma_{e_1\rightarrow f}^{(q,I=0)}\left[\frac{1}{3}+\left(\frac{2}{3}-|M_{J_{e_1}}|\right)g^{(2)}_{ave} \right]\\
   &\sigma_{e_1\rightarrow f}^{(q,I=0)}=2\pi(2\pi\alpha\omega)^{q}\left|T_{f,e_1}^{(q)}\right|^2, \qquad  M_{J_{e_1}}=0,\pm1
\end{split}
\end{equation}
where $g^{(2)}_{ave}=0.8644$, which is obtained by averaging over the isotope nuclear spins and in the long time limit. 
Eq. \ref{eq2-3} gives the formula to compute partial cross-sections in Fig. \ref{3}.

\nocite{*}

\bibliographystyle{unsrt}
\bibliography{Ba}

\end{document}